\documentclass[portrait,a0,final]{a0poster} %landscape, final
\usepackage[dvips]{epsfig}
\usepackage[dvips]{color}
\usepackage{german}
\usepackage[absolute, overlay]{textpos}
\usepackage{pst-all}      % From PSTricks
\usepackage{pst-poly}     % From pstricks/contrib/pst-poly
\usepackage{amsfonts, amsmath, amscd, amssymb}  
\usepackage{graphicx,psfrag} 

\textblockorigin{0mm}{0mm}

\definecolor{heaven}{rgb}{0.32,0.04,0.95}
\definecolor{horizon}{rgb}{0.39,0.70,0.95}
\definecolor{captionbar}{rgb}{0.80,0.77,0.38}
\definecolor{boxbackgr}{rgb}{0.9,0.9,0.9}

\definecolor{bkgr1}{rgb}{0.95,0.95,0.80}
\definecolor{bkgrheader}{rgb}{0.55,0.95,0.90}
\definecolor{hellgrau}{rgb}{0.8,0.8,0.8}
\definecolor{weiss}{rgb}{1,1,1}
\definecolor{titelhg}{hsb}{0.723,1.000,1.000}
\definecolor{titel1}{hsb}{0.167,1.000,1.000}
\definecolor{titel2}{hsb}{0.120,0.776,1.000}
\definecolor{ueberschriften}{hsb}{0.000,1.000,0.7}
\definecolor{parameterueberschrift}{hsb}{0.100,1.000,0.663}
\definecolor{params}{hsb}{0.100,1.000,0.163}
\definecolor{Gold}{rgb}{1.,0.5,0.}
\definecolor{mygold}{rgb}{1.,0.714,0.428}%{1.,0.622,0.244}
\definecolor{myyellow}{rgb}{1.0,1.0,0.5}%0.244
\definecolor{myred}{rgb}{1.0,0.465,0.465}%0.293
\definecolor{myblue}{rgb}{0.4,0.4,1.0}
\definecolor{mylightblue}{rgb}{0.75,0.75,1.0}

\usepackage{enumerate}
\usepackage{multirow}
\usepackage{amssymb,amsmath}
\usepackage[latin1]{inputenc}
\usepackage{psfrag}
\usepackage{version}
\usepackage{color}
\usepackage{graphicx}
\usepackage{multicol}

\newcommand{\kommentar}[1]{}
\newcommand{\bdm}{\begin{displaymath}}
  \newcommand{\edm}{\end{displaymath}}
\newcommand{\bequ}{\begin{equation}}
  \newcommand{\eequ}{\end{equation}}
\newcommand{\bequn}{\begin{equation*}}
  \newcommand{\eequn}{\end{equation*}}
\newcommand{\bea}{\begin{eqnarray}}
  \newcommand{\eea}{\end{eqnarray}}
\newcommand{\bean}{\begin{eqnarray*}}
  \newcommand{\eean}{\end{eqnarray*}}
\newcommand{\bi}{\begin{itemize}}
  \newcommand{\ei}{\end{itemize}}

\newcommand{\mycaption}{\psshadowbox[fillstyle=solid,
  fillcolor=captionbar, shadowsize=0.5cm]}

\newcommand{\mybox}{\psshadowbox[fillstyle=solid,
  fillcolor=weiss, shadowsize=0.5cm]}

\newcommand{\myheaderbox}{\psshadowbox[fillstyle=solid,
  fillcolor=captionbar, shadowsize=0.5cm]}

\newcommand{\mybackgndbox}{\psshadowbox[fillstyle=gradient,
  gradmidpoint=1, gradangle=0, gradbegin=heaven,
  gradend=horizon, framearc=0, shadowsize=0]}

\include{define}

\begin{document}

% ----------------------------- HEADER -----------------------------------

\begin{textblock}{74}(4,5)
  \myheaderbox
  {    
    \begin{minipage}[t][210mm][t]{740mm}
      \begin{center}
        \begin{minipage}[  ][12cm][c]{12cm}
          \centerline{\epsfig{file=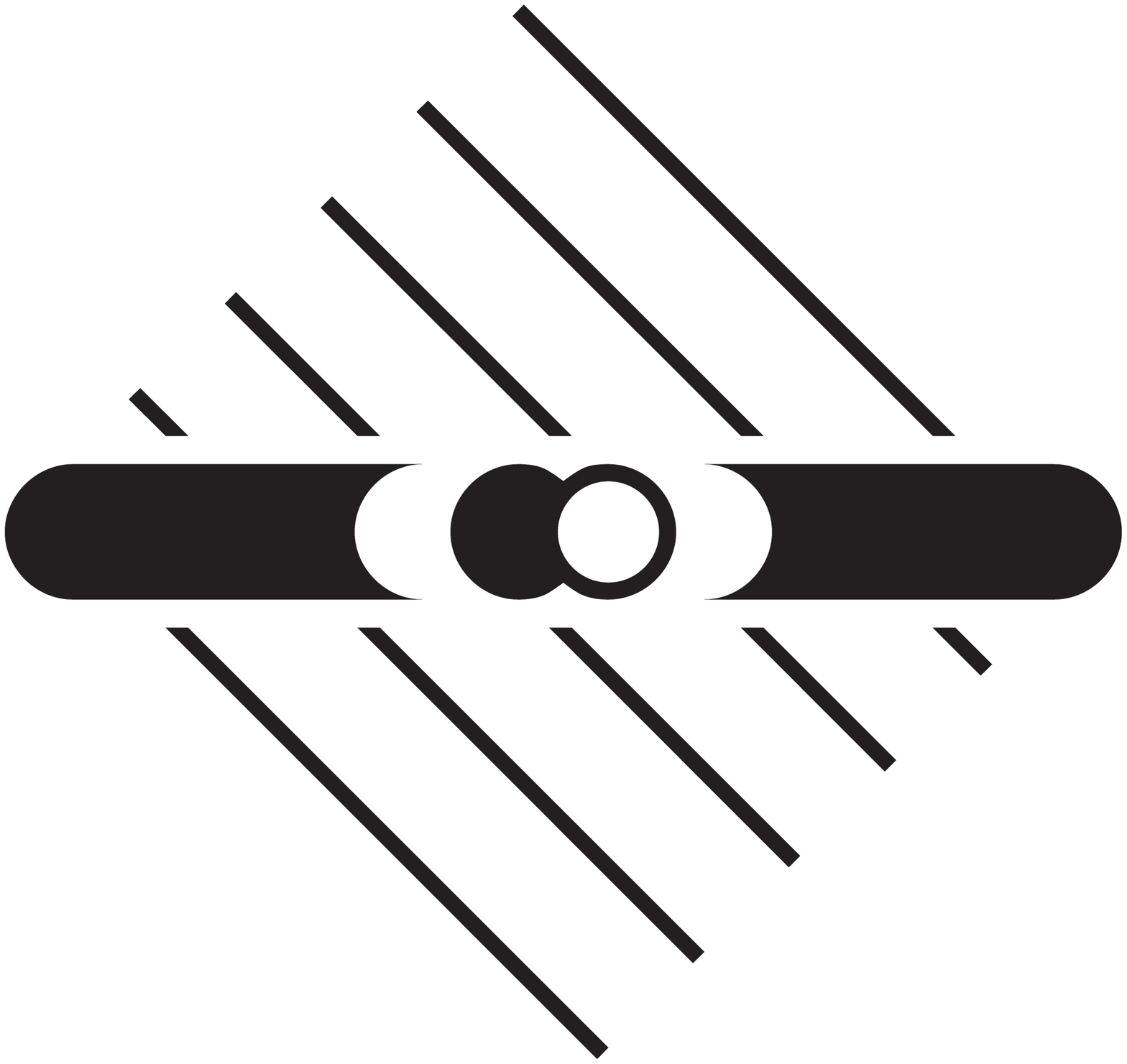,width=12cm}}
        \end{minipage}
        \begin{minipage}[][20cm][c]{456mm}
          \begin{center}
            \vspace{1cm}
            \Huge \bf Enhanced relativistic High-Harmonic Generation (HHG) using tailored pulses
            \vspace{0.7cm}
            \huge \sc Michael Klaiber$^{1,2}$, Karen Z. Hatsagortsyan$^1$,\\
            and Christoph H. Keitel$^1$
            $\;$\\[5mm]
            \huge \sl $^1$Max-Planck-Institut f\"ur Kernphysik\\
            \sl Heidelberg, Germany\\
            \huge \sl $^2$Theoretische Quantendynamik\\
            \sl Universit\"at Freiburg, Germany
          \end{center}
        \end{minipage}
        \begin{minipage}[][11cm][c]{11cm}
          \includegraphics[width=12cm]{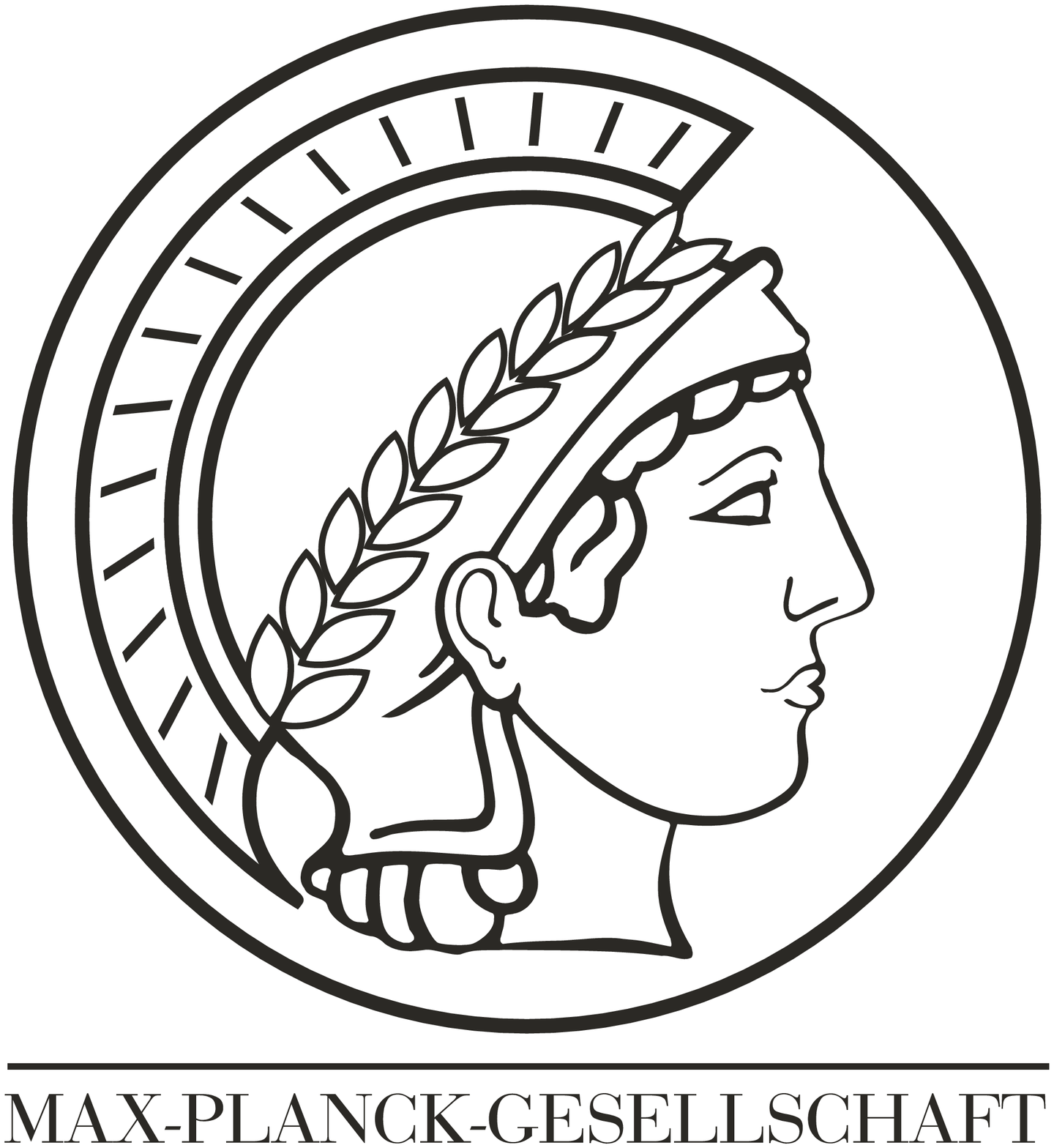}
        \end{minipage}
      \end{center}
    \end{minipage}
  }
\end{textblock}

% ----------------------------- CAPTIONS ----------------------------------

\begin{textblock}{35}(4,28)
  \mycaption
  {
    \begin{minipage}[t][30mm][t]{350mm}
      $\;$\\[8mm]
      \centerline{\Large \bf 1. Problem}    
    \end{minipage}
  } 
\end{textblock}

\begin{textblock}{35}(4,67)
  \mycaption
  {
    \begin{minipage}[t][30mm][t]{350mm}
      $\;$\\[8mm]
      \centerline{\Large \bf 2. Idea}    
    \end{minipage}
  } 
\end{textblock}

\begin{textblock}{35}(4,84)
  \mycaption
  {            
    \begin{minipage}[t][30mm][t]{350mm}
      $\;$\\[8mm]
      \centerline{\Large \bf 3. Theory}
    \end{minipage}
  }	      
\end{textblock}

\begin{textblock}{35}(43,28)
  \mycaption
  {            
    \begin{minipage}[t][30mm][t]{350mm}
      $\;$\\[8mm]
      \centerline{\Large \bf 4. The optimized tailored pulse}
    \end{minipage}
  }	      
\end{textblock}

\begin{textblock}{35}(43,65)
  \mycaption
  {            
    \begin{minipage}[t][30mm][t]{350mm}
      $\;$\\[8mm]
      \centerline{\Large \bf 5. Spectra}
    \end{minipage}
  }	      
\end{textblock}

% ----------------------------- PAGE1 -------------------------------------

\begin{textblock}{35}(4,32)      
  \mybox
  {
    \begin{minipage}[t][340mm][t]{350mm}
      $\;$\\[5mm]
      \hspace*{1.5cm}
      \parbox{32cm}
      {
        $\;$\\[1mm]  
        \small
        \vspace{-0.4cm}
        {\bf The HHG process in an atomic system:}
        \vspace{-0.4cm}\\
        \small
        \bi
      \item Ionization of a bound electron
      \item Propagation of the ionized electron in the continuum $\rightarrow $ energy gain
      \item Recombination of the ionized electron with the atomic core $\rightarrow$ Emission of high-harmonics   
        \ei       	 
        \small
        $\rightarrow$ {\color{red} {\bf the HHG process is a potential source of highly energetic radiation}}\vspace{-0.5cm}
        \\
        $\;$\\[5mm]
        {\bf Parameters:}
        \bi
        \small
      \item optical laser wavelength $\lambda =456$ nm (angular
        frequency $\omega =0.1$ a.u.)
      \item atomic ionization potential $I_p$ at the borderline between tunneling and over-the-barrier ionization
      \item laser intensities up to $5\times 10^{19}$ W/cm$^2$, i.e. strongly relativistic
        \vspace{-0.2cm}
        \\
        \ei
        \vspace{-2cm} 
        \hspace*{-1.3cm}
        \begin{minipage}[t][155mm][t]{350mm}
          \begin{center}
            \begin{minipage}[  ][13cm][c]{16cm}
              \small
              {\bf The nonrelativistic regime [1]:}
              \bi
              \small
            \item $v/c\ll 1$, laser magnetic field neglectable, no drift of the ionized electron in laser propagation direction
            \item effective recombination of the ionized electron with the atomic core $\rightarrow$ high emission rate of HHG
            \item nonrelativistic description valid up to intensities of $10^{17}$ W/cm$^2$
            \item maximal energy of the emitted radiation: $3U_p+I_p=6$ keV
              \ei
            \end{minipage}
            \begin{minipage}[][14cm][c]{160mm}
              \begin{center}
                % $\;$\\[45mm]
                \includegraphics[width=12cm]{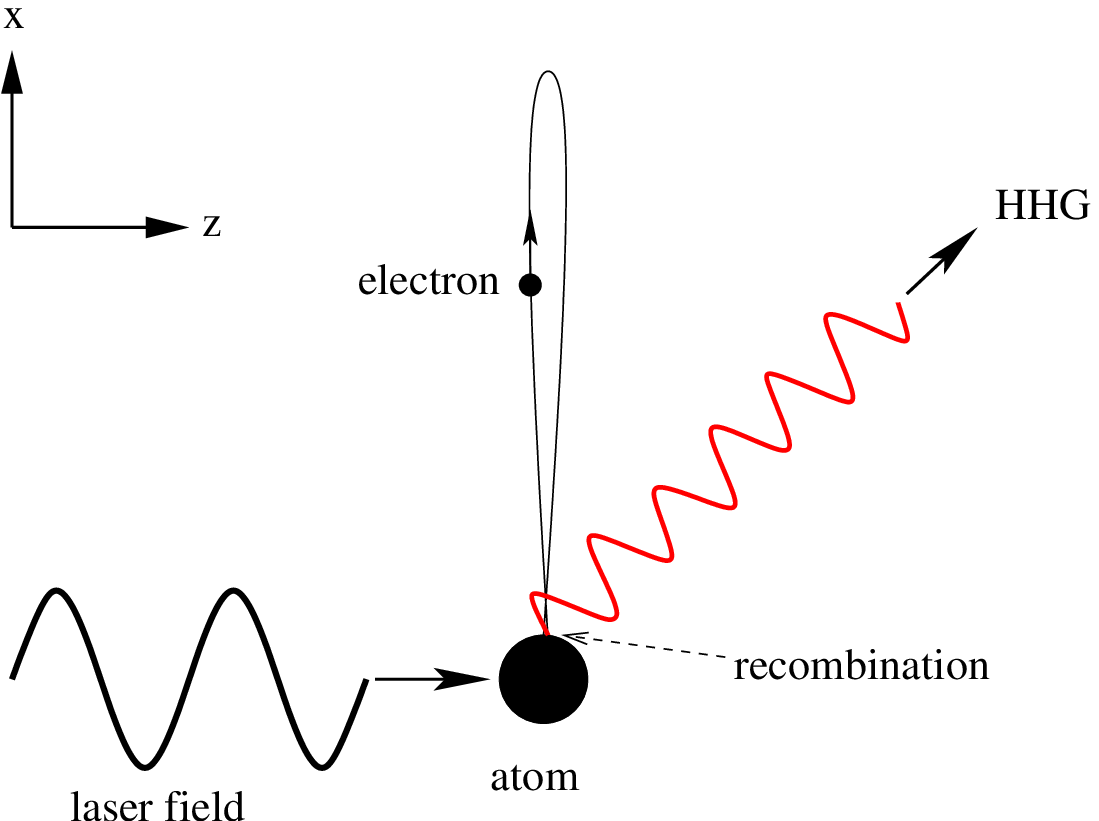}
              \end{center}
            \end{minipage}
          \end{center}
        \end{minipage}
        \vspace{-4.1cm}
        \\
        \hspace*{-1.3cm}
        \begin{minipage}[t][155mm][t]{350mm}
          \begin{center}
            \begin{minipage}[  ][13cm][c]{16cm}
              \small
              {\bf The relativistic regime [2]:}
              \bi
              \small
            \item laser magnetic field nonnegligible, drift of the ionized electron in laser propagation direction $\rightarrow$ 
              most of the electron wave packet misses the atomic core
            \item emission rate strongly suppressed
              \ei
              \small
              $\rightarrow$
              {\color{red} {\bf It is not possible to generate highly
                  energetic radiation with the HHG process in an atomic system 
                  by a simple increase of the laser intensity}}
            \end{minipage}
            \begin{minipage}[][14cm][c]{160mm}
              \begin{center}
                % $\;$\\[45mm]
                \includegraphics[width=12cm]{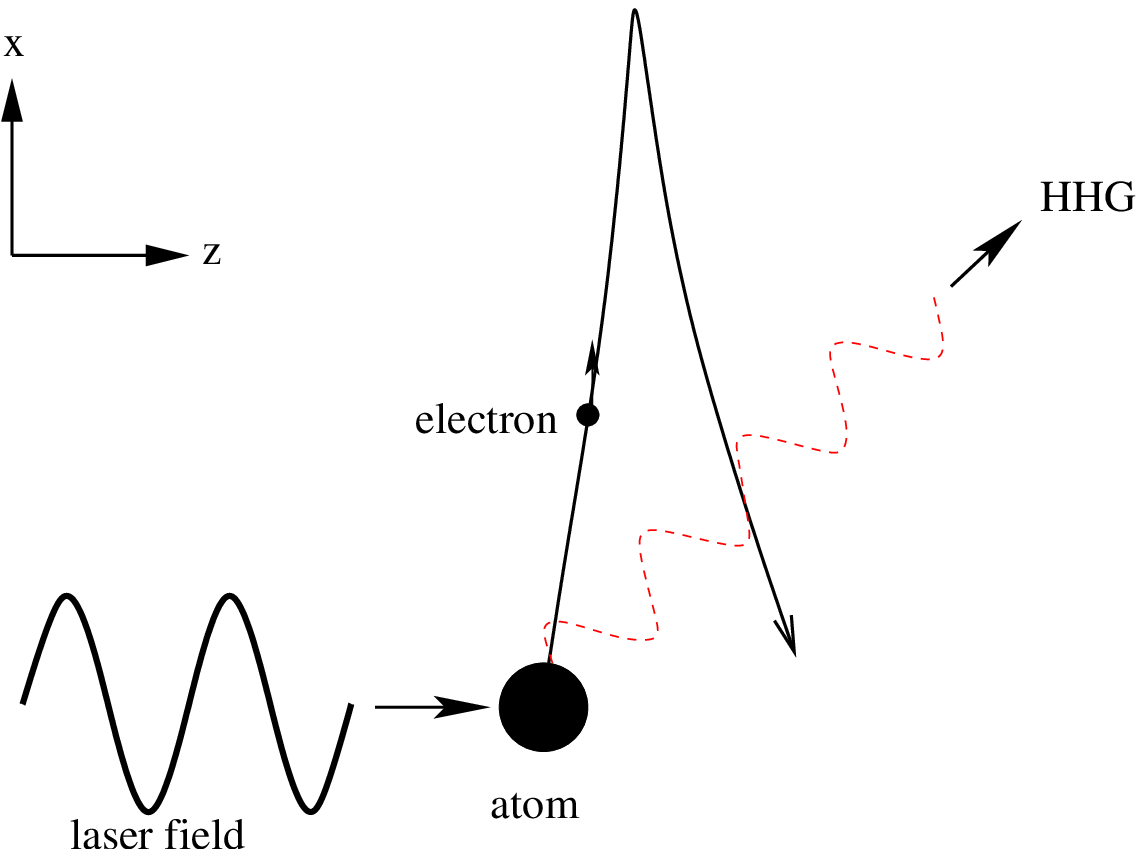}
              \end{center}
            \end{minipage}
          \end{center}
        \end{minipage}
        
      }
    \end{minipage}
  }
\end{textblock}

%%%%%%%%%%%%%%%%%%%%%%%%%%%%%%%%%%% Page2%%%%%%%%%%%%%%%%%%%%%%%%%%

\begin{textblock}{35}(4,71)      
  \mybox
  {
    \begin{minipage}[t][120mm][t]{350mm}
      $\;$\\[5mm]
      \hspace*{1.5cm}
      \parbox{32cm}
      {
        \vspace{-0.1cm}
        \small
        {\bf Is it possible to reduce the drift of the ionized electron significantly by changing the \\
          shape of the laser field from sinusoidal oscillations to
          specially tailored laser pulses?}
        \vspace{-0.1cm}      
        \\
        \begin{center}
          \psfrag{xlabel}[c]{\tiny wavenumber [$ \mbox{cm}^{-1}$]}
          \psfrag{ylabel}[b]{\tiny absorption [arb.u.]}
          \psfrag{Monomer} {\tiny \hspace{-0.2cm}monomer}
          \psfrag{Aggregat} {\tiny \hspace{-0.2cm}aggregate}
          \includegraphics[width=0.30\textwidth]{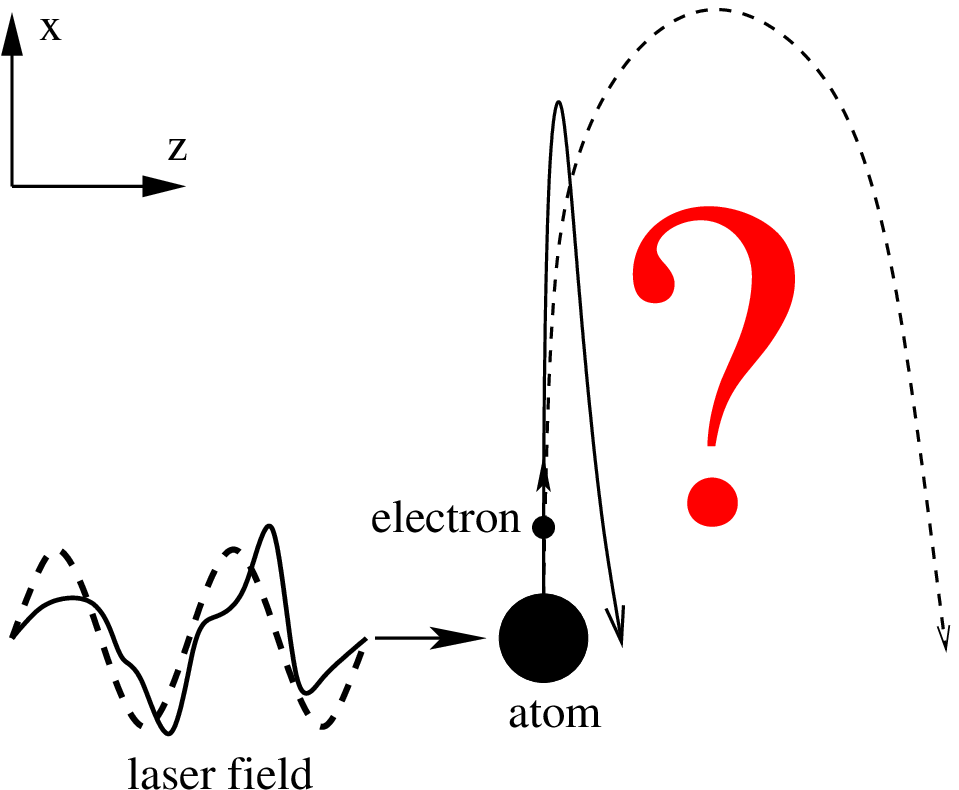}
        \end{center}
      }
    \end{minipage}
  }
\end{textblock}

%%%%%%%%%%%%%%%%%%%%%%%%%%%%%%%%%% Page3%%%%%%%%%%%%%%%%%%%%%%%%%

\begin{textblock}{35}(4,88)      
  \mybox
  {
    \begin{minipage}[t][250mm][t]{350mm}
      $\;$\\[5mm]
      \hspace*{1.5cm}
      \parbox{32cm}
      {
        \vspace{-0.2cm}
        \bi
        \small
      \item neglecting the spin of the ionized electron its dynamics is described by the Klein-Gordon equation
      \item applying the single-active electron approximation
        \vspace{-0.3cm}
        \begin{eqnarray}
          \left\{\left(i\partial^{\mu}+A^{\mu}/c+A^\mu_H/c+g^{\mu 0}V/c\right)^2
            \nonumber
            -c^2\right\}
          \Psi(x)=0
          \label{kg}
        \end{eqnarray}
        \small
        
        \ei 
        \small
        {\bf Amplitude of the HHG process within the strong-field approximation (SFA) [3]:}
        
        \vspace{-1cm}
        \small 
        
        \begin{eqnarray}
          M_{n}&=&
          -i \int
          d^4x'\int d^4x''\, \Phi(x')^*
          V_{H}(x^{\prime})G^V(x^{\prime},x^{\prime\prime})
          V_{A}(x^{\prime\prime})\Phi(x^{\prime\prime})\nonumber\\
          &=&\int^
          {\infty}_{-\infty}d\eta^{\prime}\int^{\eta^{\prime}}_{-\infty}d\eta^{\prime \prime}\int d^3\mathbf{q}\,
          m^H(\mathbf{q},\eta',\eta'') \,
          \exp\left\{-i\left[S(\mathbf{q},\eta^{\prime},\eta'')-n\eta^{\prime}\right]\right\}
          \nonumber
        \end{eqnarray}
        \bi
        \small
      \item $V_H(x)=2\mathbf{A}_H(x)\cdot(\hat{\mathbf{p}}+\mathbf{A}(\eta)/c)$ part of the electron-field interaction operator
      \item $V_{A}(x)=2iV/c^2\partial_{t}+V^2/c^2$ electron-ion interaction operator

      \item $S(\mathbf{p},\eta,\eta^{\prime})= \int^{\eta}_{\eta^{\prime}}
        d\tilde{\eta}\left(\tilde{\varepsilon}_{\mathbf{p}}
          (\tilde{\eta})-c^2+I_p\right)/\omega$ the quasiclassical action
        \ei
        \vspace{0.5cm}
        {\bf Differential rate of HHG with emission angle $\Omega$:}
        \vspace{-0.3cm}
        \small
        \begin{eqnarray} 
          \frac{dw_n}{d\Omega}=\frac{\omega^2}{(2\pi)^3}\frac{n\omega}{c^3}|M_n|^2
          \nonumber
        \end{eqnarray}
        
        {\bf Saddle-point method:} $\partial_i (S(q,\eta',\eta'')-n\eta')=0$
        \bi
        \small 
      \item valid in the long wavelength regime: $\omega<I_p$
        
        % \small
        % \vspace{-0.6cm}
        % \begin{eqnarray}
        %   \tilde{q}_x&=&
        %   -\frac{\int^{\tilde{\eta}'}_{\tilde{\eta}''}d\tilde{\eta}
        %     A(\tilde{\eta})/c}{\tilde{\eta}'-\tilde{\eta}''}
        %   \nonumber\\
        %   \tilde{q}_y&=&0
        %   \nonumber\\
        %   \tilde{q}_z&=&\frac{\tilde{q}_x^2+q_m^2/2}{\sqrt{c^2-q_m^2-\tilde{q}_x^2}}
        %   \nonumber\\
        %   \tilde{\varepsilon}_{\tilde{\mathbf{q}}}(\eta^{\prime})&=&c^2-I_p+n\omega
        %   \nonumber\\
        %   \tilde{\varepsilon}_{\tilde{\mathbf{q}}}(\eta^{\prime \prime})&=&c^2-I_p
        %   \nonumber
        %   frac{d\tilde{\varepsilon}_{\tilde{\mathbf{q}}}(\eta^{\prime})}{d \eta^{\prime
        %       prime}}&=&0,
        % \end{eqnarray}
        % \vspace{-1.3cm}
      \item defines the  recombination phase $\Re({\eta'})$, the ionization phase $\Re({\eta''})$ and the drift momentum $\tilde{\mathbf{q}}$ of the ionized electron
      \item $\Im({\eta''})$ corresponds to the ionization probability
        
        \ei

      }
    \end{minipage}
  }
\end{textblock}

% --------------------------------Page4----------------------------

\begin{textblock}{35}(43,32)
  \mybox	
  {
    \begin{minipage}[t][320mm][t]{350mm}
      $\;$\\[10mm]
      \hspace*{15mm}
      \parbox{32cm}
      {
        \vspace{-0.3cm}
        \small
        {\bf Construction of the optimized pulse [4]:}
        \bi
        \small
        \vspace{-0.2cm}
      \item Analysis of the saddle points with the Fourier expansion of an arbitrary pulse $E(\eta)=\sum^K_{k=-K} c_k \exp(ik\omega_0 \eta)$
      \item Variation of the coefficients $c_k$
      \item Minimizing $\Im({\eta''})$
        \ei
        \small
        $\rightarrow$ optimized pulse
        \vspace*{7mm}
        \\
        {\bf Features of the optimized pulse:}
        \bi 
        \small
      \item strong decrease of the laser field strength after the phase of ionization\\
        $\rightarrow$ concentration of the ionization force on a short time interval
        \vspace{0.2cm}
      \item long time span with weak laser field strength after ionization\\
        $\rightarrow$ Propagation with nonrelativistic velocity, i.e. no drift
        \vspace{0.2cm}
      \item  strong and short pulse shortly before recombination\\
        $\rightarrow$ concentration of the acceleration force on a short time interval, i.e. fast energy gain, little drift\\ 
        \ei 
        \vspace{-0.4cm}
        {\bf The optimized pulse:}
        \bi
        \small
      \item Synopsis of the discovered features plus simplification\\
        $\rightarrow$ attosecond pulse train with rectangular pulses with a duration of 90 as
        (laser period 1.5 fs)\\
        \ei 
        \begin{center} 
          \includegraphics[width=12cm,clip=true]{pulse.eps}
          \includegraphics[width=12cm,clip=true]{abst.eps}
        \end{center}
        \small
        (i) Temporal dependence of the tailored pulse compared with a sinus wave with the same averaged intensity.\\ 
        (ii) The frequency spectrum of the tailored pulse. The phase spectrum of the tailored pulse is linear, i.e. the phases are locked.
          
      }

    \end{minipage}
  }
\end{textblock}

% ------------------------Page5-----------------------------------------

\begin{textblock}{35}(43,69)
  
  \mybox	
  {
    \begin{minipage}[t][340mm][t]{350mm}
      $\;$\\[10mm]
      \hspace*{15mm}
      \parbox{32cm}
      {
        \vspace{-0.3cm}
        \small
        (i)  moderately relativistic regime {\bf $\rightarrow$ maximal harmonic emission energy of 100 keV }\\
        (ii) strongly relativistic regime {\bf $\rightarrow$ maximal harmonic emission energy of 1 MeV }

        \begin{center}
          \includegraphics[width=14cm,clip=true]{h5x18.eps} 
          \includegraphics[width=14cm,clip=true]{h5x19.eps}
        \end{center}
        \small
        \begin{center}Harmonic emission rate in laser polarization direction via $\log_{10}(dw_n/d\Omega)$,
          as function of the harmonic energy with a laser intensity of (i) $5\times 10^{18}$ W/cm$^2$
          and (ii) $5\times 10^{19}$ W/cm$^2$. The  main angular frequency
          is $\omega=0.1$ a.u.. The ionization potential is (i) $I_p=28$ a.u. and (ii) $I_p=62$ a.u.,
          respectively: (a, grey) within the dipole approximation and a sinusoidal field,
          (b, dashed) with respect to the Klein-Gordon
          equation and a sinusoidal field, (c, black) with respect to the Klein-Gordon
          equation and a tailored pulse.
        \end{center}
        \vspace{0.5cm}
        % \\
        \begin{center}
          \includegraphics[width=14cm,clip=true]{h120.eps}
        \end{center}
        \small
        \begin{center}Harmonic emission spectra in laser polarization direction via $\log_{10}(dw_n/d\Omega)$,
          as function of the harmonic energy with a laser intensity of $5\times 10^{19}$ W/cm$^2$ using the tailored pulse, which consists of 20, 40, 60, 80, 100, 
          and 120 harmonics of the main angular frequency. The ionization potential is $I_p=62$ a.u..
        \end{center}
        
      }

    \end{minipage}
  }
\end{textblock}

% ----------------------------- FOOTER ------------------------------------

\begin{textblock}{74}(43,104)
  \mybox
  {
    \begin{minipage}[t][90mm][t]{350mm}
      $\;$\\[10mm]
      \hspace*{15mm}
      \parbox{35cm}
      {
        [1] M. Lewenstein, P. Balcou, M.Yu. Ivanov, A. Huillier, and P.B.~Corkum,\\ 
        \hspace*{10mm} Phys. Rev. A {\bf 49}, 2117 (1994).\\ $\;$
        [2] M.W.~Walser, C.H. Keitel, A. Scrinzi, and T. Brabec,\\ 
        \hspace*{10mm} Phys. Rev. Lett. {\bf 85}, 5082 (2000).\\ $\;$
        [3] D. B. Milosevic, S. X. Hu, and W. Becker, \\ 
        \hspace*{10mm}Phys. Rev. A {\bf 63}, 011403 (2000).\\$\;$
        [4] M. Klaiber, K.Z. Hatsagortsyan, and C.H. Keitel, to be published (2006).
        
      }

    \end{minipage}
  }
\end{textblock}

% ------------------------------BACKGROUND --------------------------------

\begin{textblock}{78}(2,3)
  \mybackgndbox
  {
    \begin{minipage}[t][1120mm][t]{780mm}
      $\;$
    \end{minipage}
  }
\end{textblock}
% ----------------------------- END OF DOCUMENT ---------------------------
\end{document}